\documentclass[twocolumn,preprintnumbers,nofootinbib,prd]{revtex4}
\usepackage{psfrag,graphicx,epsfig,amsmath,amssymb,bm}
\def\beq{\begin{equation}}
\def\eeq{\end{equation}}
\def\beqa{\begin{eqnarray}}
\def\eeqa{\end{eqnarray}}
\def\eqn#1{eq.~(\ref{#1})}
\def\Eqn#1{Equation~(\ref{#1})}
\def\eqns#1#2{eqs.~(\ref{#1}) and~(\ref{#2})}

\begin{document}

\preprint{IPPP/11/47 -- DCPT/11/94 -- Edinburgh 2011/21 -- DFTT-15/2011} 
%
% arXiv:0901.nnnn [hep-ph]

\title{\boldmath
An infrared approach to Reggeization
\unboldmath}

\author{Vittorio Del Duca\,$^a$, Claude Duhr\,$^b$, 
Einan Gardi\,$^c$, Lorenzo Magnea\,$^d$ and Chris D. White\,$^e$}

\affiliation{$^a$ INFN, Laboratori Nazionali Frascati, 00044 
Frascati (Roma), Italy\\ 
$^b$ Institute for Particle Physics Phenomenology, University 
of Durham, Durham, DH1 3LE, UK\\ 
$^c$ The Tait Institute, School of Physics and Astronomy, 
The University of Edinburgh, Edinburgh EH9 3JZ, Scotland, UK \\ 
$^d$ Dipartimento di Fisica Teorica, Universit{\`a} di Torino, 
and INFN, Sezione di Torino, Via P. Giuria 1, I-10125 Torino, Italy \\
$^e$ School of Physics and Astronomy, Scottish Universities Physics 
Alliance, University of Glasgow, Glasgow G12 8QQ, Scotland, UK.}

\begin{abstract}

\noindent
We present a new approach to Reggeization of gauge amplitudes 
based on the universal properties of their infrared singularities. 
Using the ``dipole formula'', a compact ansatz for all infrared 
singularities of massless amplitudes, we study Reggeization of 
singular contributions to high-energy amplitudes for arbitrary 
color representations, and any logarithmic accuracy. We derive 
leading-logarithmic Reggeization for general cross-channel color 
exchanges, and we show that Reggeization breaks down 
for the imaginary part of the amplitude at next-to-leading 
logarithms and for the real part at next-to-next-to-leading 
logarithms. Our formalism applies to multiparticle amplitudes 
in multi-Regge kinematics, and constrains possible corrections 
to the dipole formula starting at three loops.

\end{abstract}

%\pacs{11.15.Bt,12.38.Bx,12.38.Cy,13.87.-a}

\maketitle

%CITE DM EARLIER
\noindent
{\bf Introduction}. 
The high-energy limit of gauge theory scattering amplitudes is
of great interest both from a theoretical standpoint and in
view of phenomenological applications. It has, therefore, been
studied in depth for many years, starting with work done before 
the development of the Standard Model of particle 
physics~\cite{ELP}. The key feature of the high-energy limit 
is the phenomenon of Reggeization, which can be economically 
described in the simple case of four-point massless gauge theory 
amplitudes, characterized by the Mandelstam invariants $s$ (the 
center-of-mass energy), $t$ and $u$, satisfying $s + t + u = 0$. 
In the high-energy limit, $|s/t| \to \infty$, amplitudes which are 
dominated at the lowest perturbative order by the exchange of a 
given state in the $t$ channel, are found to receive logarithmic 
corrections in powers of $\ln (s/(-t))$. These corrections can 
be resummed to all orders in perturbation theory by the simple 
prescription of replacing the tree-level $t$-channel propagator 
according to
\beq
  \frac{1}{t} \, \longrightarrow \, \frac{1}{t} \, 
  \left( \frac{s}{- t} \right)^{\alpha(t)} \, ,
\label{propregg}
\eeq
where $\alpha(t)$ is the Regge trajectory, which can be expanded
in powers of the gauge coupling $\alpha_s$. The Regge trajectory
is infrared divergent, since its formal definition involves integrations
over the transverse momentum of virtual gauge bosons. It is therefore
appropriate to resort to dimensional regularization, and express the
trajectory in terms of the $d$-dimensional coupling (with $d = 4 - 2 
\epsilon$, $\epsilon <0$), evaluated at the scale $\mu^2 = - t$. 
One then writes 
\beq
  \alpha (t) = \frac{\alpha_s (-t, \epsilon)}{4 \pi} \, \alpha^{(1)}
  \, + \left( \frac{\alpha_s (-t, \epsilon)}{4 \pi} \right)^2
  \alpha^{(2)} + {\cal O} \left( \alpha_s^3 \right) \, ,
\label{pertreg}
\eeq
where $\alpha_s(- t, \epsilon) = (\mu^2/(-t))^{- \epsilon} 
\alpha_s (\mu^2) + {\cal O} (\alpha_s^2)$. The replacement rule 
in \eqn{propregg} can be generalized to the case of $2 \to n$ 
amplitudes in the so-called `multi-Regge' kinematic (MRK) 
regime~\cite{Kuraev:1976ge,DelDuca:1995hf}. When the $n$ 
emitted partons are strongly ordered in rapidity, but have comparable 
transverse momenta, if the tree-level amplitude is dominated by 
the $t$-channel exchange of a given color state, then leading 
logarithmic (LL) virtual corrections to the amplitude are resummed 
to all orders by replacing the $t$-channel propagator between the 
emission of particle $k$ and particle $k+1$ according to
\beq
  \frac{1}{t_k} \longrightarrow \frac{1}{t_k} \,
  {\rm e}^{\alpha (t) \, (y_k - y_{k+1})}
  \equiv \frac{1}{t_k} \left(- \frac{s_{k, k+1}}{t_k} 
  \right)^{\alpha (t)} \, ,
\label{propdress}
\eeq
where $y_k$ are the rapidities of the emitted particles, and
$s_{k, k+1} \equiv (p_k + p_{k + 1})^2 = 2 p_k \cdot 
p_{k + 1}$. Reggeization has been proved for $t$-channel
gluon exchange at LL~\cite{Balitsky:1979ap} and NLL 
accuracy~\cite{Fadin:2006bj}, and for quark exchange at 
LL accuracy~\cite{Bogdan:2006af}. Furthermore, the 
quark~\cite{Bogdan:2002sr} and gluon~\cite{Fadin,Blumlein:1998ib,DelDuca:2001gu} 
Regge trajectories have been determined up to two loops.

Reggeization proofs typically rely upon intricate recursive
arguments, and depend upon the specific state that dominates 
$t$-channel exchanges for the process at hand. We propose a 
different viewpoint, based upon recent advances in our understanding 
of the all-order structure of infrared divergences in massless 
gauge theories. This approach is made possible~\cite{Korchemsky:1993hr,Dokshitzer:2005ig} by the fact that the Regge trajectory $\alpha(t)$ is infrared divergent, and indeed is largely 
determined by infrared singularities (non-trivial finite contributions start arising only at two loops). Furthermore, singular terms have 
been empirically shown to be universal: for quarks and gluons at 
one loop one finds that $\alpha^{(1)} = 2 \, C_R/\epsilon$, where 
$C_R$ is the Casimir of the appropriate color representation, 
$C_A$ for gluons and $C_F$ for quarks. This suggests that 
Reggeization can be efficiently and  generally studied from an 
infrared point of view.

\noindent
{\bf The infrared approach}.
Our main tool is the `dipole formula'~\cite{Becher:2009cu,
Gardi:2009qi,Becher:2009qa}, an all-order ansatz for infrared 
divergences of fixed-angle massless gauge theory amplitudes. 
Such amplitudes can be expressed in the factorized form
\beq
  {\cal M} \left(\left\{p_i\right\}, \alpha_s,  \epsilon \right)  
  \, = \, Z \left(\left\{p_i\right\}, \alpha_s,  \epsilon \right) 
  \, \, {\cal H} \left(\left\{p_i\right\}, \alpha_s, \epsilon \right) \, ,
\label{Mfac}
\eeq
where $p_i$, $i = 1, \ldots, L$, are the hard parton momenta,
${\cal H}$ is the hard part of the amplitude, finite as $\epsilon 
\to 0$, and $Z$ is the operator responsible for all infrared and 
collinear divergences. The dipole formula is a compact expression 
for the $Z$ operator, stating that only two-parton correlations 
appear at the level of the exponent. In the color generator notation~\cite{Catani:1996vz}, one finds
\beqa
\label{sumodipoles}
  && Z \left(\left\{p_i\right\}, \alpha_s, \epsilon \right) \, = \, 
  \exp \Bigg\{ \int_0^{\mu^2} \frac{d \lambda^2}{\lambda^2}
  \bigg[\frac{1}{4} \widehat{\gamma}_K \left( \alpha_s(
  \lambda^2, \epsilon) \right) \,\times
  \nonumber \\ & & 
  \sum_{i<j}  \ln \left(\frac{-s_{ij}}{\lambda^2} \right)
  {\bf T}_i \cdot {\bf T}_j - \frac{1}{2} \sum_{i = 1}^L
  \gamma_i \left(\alpha_s (\lambda^2, \epsilon)
  \right) \bigg] \Bigg\} \, .
\eeqa
Here $- s_{ij} = 2 \left\vert p_i \cdot p_j \right \vert e^{ - {\rm i}
\pi \lambda_{ij}}$, with $\lambda_{i j} = 1$ if partons $i$ and $j$ 
both belong to either the initial or the final state, $\lambda_{i j} = 0$ 
otherwise; ${\bf T}_i$ are color generators in the representation of 
parton $i$, acting on the color indices of the amplitude as described 
in~\cite{Catani:1996vz}; $\widehat{\gamma}_K (\alpha_s)$ is the 
cusp anomalous dimension~\cite{Korchemsky:1985xj}, with the 
Casimir of the appropriate representation scaled out; $\gamma_i$ 
are the anomalous dimensions of the fields associated with external 
particles. 

The dipole formula~\cite{Gardi:2009qi,Becher:2009qa} is the
simplest solution to a set of exact equations governing infrared
singularities for massless particles, which in turn follow from
factorization properties of fixed-angle matrix elements in soft 
and collinear limits, and from rescaling invariance of light-like Wilson 
lines. The dipole formula is known to be exact up to two loops
in the exponent~\cite{Aybat:2006mz}, and corrections can only 
arise, starting at three loops, in the form of highly constrained 
functions of conformal invariant cross-ratios of external 
momenta~\cite{Gardi:2009qi}, studied in~\cite{Becher:2009qa,
Dixon:2009ur}, or, at even higher orders, if the cusp anomalous 
dimension receives contributions from higher-order Casimir 
operators~\cite{Gardi:2009qi}.

In the present context, it is important to note that the fixed-angle
assumption, which underlies \eqn{Mfac}, and which amounts to the
requirement that all kinematic invariants $s_{ij}$ be of comparable 
size, breaks down in the Regge limit $s \gg |t|$. This does not affect 
our reasoning concerning infrared singularities: indeed, all logarithms 
of $s/t$ which are accompanied by infrared poles are correctly 
captured by \eqn{Mfac}. We will, however, not be able to control 
logarithms with coefficients that remain finite as $\epsilon \to 0$, 
and thus our approach will only yield the divergent contributions to 
the Regge trajectory $\alpha (t)$.

With this important proviso, we may study the high-energy limit
of \eqn{sumodipoles}, starting with the simple but crucial case 
of the four-point amplitude. Expanding the exponent in powers 
of $t/s$, and enforcing color conservation by means of the
constraint $\sum_i {\bf T}_i = 0$, we find that to leading
power in $t/s$, and thus to \emph{any} logarithmic accuracy, the
infrared operator $Z$ can be factorized as
\beq
  Z \left( \left\{p_i\right\}, \alpha_s, \epsilon \right)
  \, = \, \widetilde{Z} \left( \frac{s}{t}, \alpha_s, \epsilon 
  \right) \, Z_{\bf 1} \left( t, \alpha_s, 
  \epsilon \right) \, ,
\label{Zfac}
\eeq
where the factor $Z_{\bf 1}$ is proportional to the identity matrix 
in color space, and is independent of $s$, while the non-trivial color 
structure and $s$ dependence are contained in the Reggeization
operator $\widetilde{Z}$. Introducing the color 
operators~\cite{Dokshitzer:2005ig} ${\bf T}_s
= {\bf T}_1 + {\bf T}_2$ and  ${\bf T}_t = {\bf T}_1 
+ {\bf T}_3$ the Reggeization operator can be compactly written as,
\beq
  \widetilde{Z} \left( \frac{s}{t}, \alpha_s, \epsilon \right) 
  = \exp \left\{ K \big(\alpha_s, \epsilon \big)
  \Bigg[ \ln \left( \frac{s}{- t} \right) {\bf T}_t^2 + 
  {\rm i} \pi  {\bf T}_s^2 \Bigg] \right\}  ,
  \label{Ztildedef}
\eeq
while the color-trivial factor $Z_{\bf 1}$ can be written as
\beqa
\label{Zddef}
  & & Z_{\bf 1} \left( t, \alpha_s, \epsilon \right)
  \, = \,\exp \left\{ \sum_{i = 1}^4 B_i \big(\alpha_s, 
  \epsilon \big) \right.  \\
  & & \left. + \nonumber
  \, \frac12 \left[K \big(\alpha_s, \epsilon 
  \big) \, \left(\ln \left(\frac{-t}{\mu^2}\right) -{\rm i}\pi\right) + 
  D \big(\alpha_s, \epsilon \big) \right] \, 
  \sum_{i = 1}^4 C_i \right\} \, ,
\eeqa
where $C_i$ are the Casimir invariants of the appropriate representations, which in the present notation are given by 
$C_i = {\bf T}_i \cdot {\bf T}_i$. In \eqns{Ztildedef}{Zddef} we 
have introduced a set of  functions encoding the dependence on 
the coupling and on $\epsilon$, through integrals over the scale 
of the $d$-dimensional running coupling. They are defined by
\beqa
  K \big(\alpha_s, \epsilon \big) & \equiv &
  - \frac14 \int_0^{\mu^2} \frac{d \lambda^2}{\lambda^2} \, 
  \widehat{\gamma}_K \left(\alpha_s(\lambda^2, \epsilon) \right) ,
  \label{Kdef} \\ 
  D \big(\alpha_s, \epsilon \big) & \equiv & 
  - \frac14 \hspace{-2pt} \int_0^{\mu^2} \hspace{-1mm}
  \frac{d \lambda^2}{\lambda^2} \, 
  \widehat{\gamma}_K \left(\alpha_s(\lambda^2, \epsilon) \right)
  \ln \left( \frac{\mu^2}{{\lambda^2}} \right) \hspace{-2pt},
  \label{Idef} \\
  B_i \big(\alpha_s, \epsilon \big) & \equiv & 
  - \frac12 \int_0^{\mu^2} \frac{d\lambda^2}{\lambda^2} \, 
  \gamma_{J_i} \left(\alpha_s (\lambda^2, \epsilon) \right) ;
  \label{Jintdef}
\eeqa
and they can be evaluated order by order in the coupling, yielding
for example
\beq
  K (\alpha_s, \epsilon) \, = \, \frac{\alpha_s}{\pi} 
  \frac{1}{2 \epsilon} + \left(\frac{\alpha_s}{\pi}\right)^2
  \left( \frac{\widehat{\gamma}_K^{(2)}}{8 \epsilon} -
  \frac{b_0}{16 \epsilon^2} \right) + {\cal O}(\alpha_s^3) \,  ,
\label{KNLO}
\eeq
where
\beq
  b_0 \, = \, {11C_A - 2 n_f \over 3}\, , \,\,\,
  \widehat{\gamma}_K^{(2)} = \left({67 \over 18} -
  {\pi^2 \over 6} \right) C_A - {5\over 9} n_f \, ,
\label{b0gam2}
\eeq
and we used the fact that $\widehat{\gamma}_K^{(1)}
= 2$ in our normalization.

\Eqn{Ztildedef} is one of our central results: it allows us 
to study Reggeization, for singular contributions to the amplitude,
on completely general grounds. In particular, at LL accuracy we observe that Reggeization is a general property of any amplitude 
which at leading order and high energy is dominated by the 
$t$-channel exchange of a state belonging to a given color 
representation. Indeed, at LL level we can neglect the imaginary 
part of the exponent of \eqn{Ztildedef}. Equation~(\ref{Mfac}) can 
then be written as
\beq
  \left.{\cal M}\right\vert_{\rm LL} \, = \,  
\left(\frac{s}{- t} \right)^{ K \,   \,
  {\bf T}_t^2 } \, Z_{\bf 1} \, {\cal H} \, .
\label{Mggdef}
\eeq
If the hard function ${\cal H}$, at tree level and at leading power 
in $t/s$, is given by the exchange of a specific color representation
$R$ in the $t$ channel, then it is an eigenstate of the $t$-channel 
operator ${\bf T}_t^2$, and LL Reggeization follows, since one can simply replace 
the operator ${\bf T}_t^2$ in (\ref{Mggdef}) by its eigenvalue $C_R$,
the Casimir operator of the appropriate representation. This implies
the universality of the LL Regge trajectory, which for any process
of the kind just described can be read off using \eqn{Mggdef},
together with \eqns{pertreg}{KNLO}, and is given by $\alpha^{(1)} 
= 2 C_R/\epsilon$. We note in passing that an identical reasoning
can be repeated for $u$-channel exchanges, in all cases in which
they are responsible for logarithmic enhancements: Reggeization
follows, with the same universal form of the Regge trajectory, given 
the symmetry of the dipole formula under $t \leftrightarrow u$. 
By the same token, it is possible to write the Reggeized amplitude 
so that it displays the correct `signature', namely the symmetry under 
$s \leftrightarrow u$ exchange.

{\bf Reggeization breaking}. We see that LL Reggeization is a 
general feature of massless gauge amplitudes, dictated by 
the universal structure of infrared divergences. The dipole ansatz (\ref{sumodipoles}), 
however, is an all-order statement on the perturbative exponent, 
allowing for a much deeper analysis. In particular, \eqn{Ztildedef} is valid 
to any logarithmic accuracy, and it can be used to study 
Reggeization beyond LL. It is immediately evident from 
\eqn{Ztildedef} that already at NLL level the simple picture of 
Reggeization given by \eqn{Mggdef} will break down for the 
imaginary part of the amplitude: indeed, if the hard 
part~${\cal H}$ is an eigenstate of the operator ${\bf T}_t^2$, 
it will generically not be an eigenstate of ${\bf T}_s^2$, which 
does not commute with~${\bf T}_t^2$. Next-to-leading logarithms 
will then mix different color structures, and a simple resummation 
of the form of \eqn{propregg} will fail. We verified that at 
NLL level only the imaginary part of the amplitude is affected, 
confirming and extending to general color exchanges the property 
of NLL Reggeization of the real part of the amplitude, with a 
universal NLO Regge trajectory given by \eqn{KNLO}, multiplied 
by the appropriate Casimir eigenvalue. Proceeding to 
NNLL~\cite{DelDuca:2001gu}, Reggeization generically breaks 
down also for the real part of the amplitude. The leading color 
operator responsible for this breakdown is
\beq
  {\cal E} \left( \frac{s}{t}, \alpha_s, 
  \epsilon \right) \equiv - \frac{\pi^2}{3}
  K^3 (\alpha_s, \epsilon)  \ln \left(\frac{s}{- t} \right) 
  \Big[{\bf T}_s^2, \big[{\bf T}_t^2, {\bf T}_s^2 \big] 
  \Big] ,
\label{NNLL}
\eeq
and it will generate a tower of non-Reggeizing logarithms, starting 
at three loops with terms of the form $- (\pi^2/3) (\alpha_s/(2 
\pi \epsilon))^3 \ln (s/(-t))$. The specific nature of the color 
mixing responsible for the breakdown of Reggeization can be 
studied case by case by applying the color operator $\left[
{\bf T}_s^2, \left[{\bf T}_t^2, {\bf T}_s^2 \right] \right]$ to 
the hard amplitude~${\cal H}$: while it is conceivable that in 
some specific cases the operator will have a vanishing eigenvalue, 
this will not happen for generic representations, and the resulting 
effect can be quantified using \eqn{NNLL}. Note that Reggeization breaking is a subleading effect in the \hbox{large-$N_c$} limit. Finally, we emphasize that although ${\cal O}(1/\epsilon)$ corrections going beyond the dipole formula (\ref{sumodipoles}) may arise at 
${\cal O}(\alpha_s^3)$~\cite{Gardi:2009qi,Becher:2009qa,Dixon:2009ur}, these cannot influence Reggeization breaking (\ref{NNLL}), which is ${\cal O}(1/\epsilon^3)$.

{\bf Constraining soft singularities.} One may also reverse the logic, 
and use what is known about the Regge limit to constrain possible 
corrections to the dipole formula. Such corrections may first 
arise at three loops. These are highly constrained even before 
considering the Regge limit: they may only depend on kinematics 
via conformally-invariant cross ratios~\cite{Gardi:2009qi} and they 
must vanish in all collinear limits~\cite{Becher:2009qa,Dixon:2009ur}. 
Nevertheless, a few explicit examples, consistent with all constraints 
were constructed~\cite{Dixon:2009ur}. Specialising these examples
to the high-energy limit, one finds that they are all characterized by 
super-leading logarithms~${\cal O}(\alpha_s^3\log^4(s/\!-t))$, 
which are inconsistent with known results on LL Reggeization. 
Thus, consistency with the Regge limit is a powerful constraint: it 
rules out all existing examples for potential three-loop corrections 
which go beyond the dipole formula. Note, however, that such corrections 
may still be present: the unphysical logs could cancel in linear combinations, 
or there might be other functions satisfying all constraints.

{\bf Extension to $2\rightarrow n$ scattering}. 
So far we have concentrated on virtual corrections to four-point
amplitudes in the high-energy limit, to illustrate the basic features
of our approach. We note however that the dipole formula applies
to amplitudes with any number of partons, and can therefore be 
used also to study Reggeization in the general case of multiparticle
amplitudes. Consider specifically a
scattering process $p_1 + p_2 \to p_3 +  \ldots + p_L$, in 
multi-Regge kinematics, where the parton rapidities are strongly ordered, $y_3 \gg y_4 \gg \ldots \gg y_L$, while their transverse momenta $k^\perp_i \simeq k^\perp_j ,
\forall i,j$ are comparable. 
In this situation the Mandelstam invariants are also
hierarchically ordered, and may be approximated by $- s_{i j} \simeq \,{\rm e}^{- {\rm i} \pi \lambda_{i j}} k^\perp_i k^\perp_j {\rm e}^{y_i - y_j}$. 
It is then possible to show that in multi-Regge 
kinematics the infrared operator $Z$ expressed by the dipole formula 
in \eqn{sumodipoles} also factorizes, generalizing \eqn{Zfac}. 
One finds
\begin{align}
  Z \left(\left\{p_i\right\} \right)
  = \widetilde{Z}^{\rm MR} \left(\!\left\{\Delta y_i\right\}\right)\, \,Z_{\bf 1}^{\rm MR} 
  \left( \left\{k_i^\perp\right\} \right) ,
\label{ZMRK} 
\end{align}
where $\Delta y_i \equiv y_i - y_{i + 1}$, and where we omitted, for simplicity, the arguments $\alpha_s$ and $\epsilon$ in the three functions. The multi-Regge 
operator $\widetilde{Z}^{\rm MR}$ and the singlet factor 
$Z_{\bf 1}^{\rm MR}$ generalize \eqns{Ztildedef}{Zddef}, respectively. 
To write them down, it is useful to define the color operators
\beq
  {\bf T}_{t_i} \, \equiv \, {\bf T}_1 + \sum_{p = 1}^i 
  {\bf T}_{p + 2} \, ,
\label{Tkdef}
\eeq
whose eigenstates are definite $t$-channel exchanges between 
partons $i+2$ and $i+3$. Using color conservation, one can 
then derive an explicit expression for 
$\widetilde{Z}^{\rm MR}$, valid for any number of emitted partons. It is given by
\begin{align}
\label{ZtildeMRK}
  \widetilde{Z}^{\rm MR} \left( \left\{\Delta y_i\right\}\right)   =  \exp \Bigg\{ K%\left( \alpha_s,   \epsilon \right) 
\,\bigg[\sum_{i = 3}^{L-1} {\bf T}^2_{t_{i - 2}}
  \, \Delta y_i 
%\nonumber \\  && 
+ \, {\rm i} \pi \, {\bf T}_s^2 \bigg] \Bigg\} \, .
\end{align}
The form of $Z_{\bf 1}^{\rm MR}$ will be given elsewhere. In
\eqn{ZMRK} the rapidity (and non-trivial colour) dependence is 
concentrated in the multi-Regge operator $\widetilde{Z}^{\rm MR}$, 
whereas the factor $Z_{\bf 1}^{\rm MR}$, which is proportional to 
the unit matrix in color space, depends only on the transverse 
momenta of the emitted partons. The requirement for Reggeization 
of the amplitude is then that the hard interaction be dominated by 
a series of $t$-channel exchanges in the Regge limit. If this is the 
case, then each $t$-channel exchange between, say, the emissions 
of partons $i + 2$ and $i + 3$, will be an eigenstate of the 
color operator ${\bf T}_{t_i}^2$, and the corresponding 
rapidity dependence will enter the exponent of the amplitude with 
the relevant eigenvalue. The simplest case is when a single particle 
species is exchanged in the $t$-channel, which effectively radiates 
all other final state particles. One then recovers the well-known 
Reggeization of leading logarithms in the form  of \eqn{propdress}. 
Reggeization, however, is more general than this, as is clear from 
\eqn{ZtildeMRK}: in principle, different $t$-channel exchanges 
may occur, so that different rapidity intervals will exponentiate 
with distinct eigenvalues. Note that, as in the four-point case, 
the $s$-channel color operator ${\bf T}_s^2$ in \eqn{ZtildeMRK} 
will generically lead to a breakdown of Reggeization for the 
imaginary part of the amplitude at NLL accuracy, and for the 
real part of the amplitude at NNLL level.

{\bf Conclusions}. We have proposed a general approach to
Reggeization of massless gauge-theory amplitudes, which uses  
the universal structure of infrared divergences embodied 
in the dipole formula, \eqn{sumodipoles}. The drawback of 
our approach is that we have no direct control of finite
contributions to the Regge trajectory. The dipole formula may also receive
corrections, starting at three loops, but we have been able to show 
that such corrections are irrelevant both for deriving Reggeization, 
and for understanding its breaking. Conversely, the Regge limit 
provides a useful constraint on potential corrections going beyond the dipole 
formula, which are already highly constrained by other means.
This strengthens the evidence for the validity of the dipole 
formula beyond two loops.
Our approach shows the complete generality of the Reggeization phenomenon at LL level, explaining
the universality of divergent contributions to the Regge 
trajectory at LL and NLL accuracy. Furthermore, we give 
compact expressions for the operators which generate 
all high-energy logarithms associated with infrared divergences, both 
for the four-point amplitude and for the multiparticle case in 
multi-Regge kinematics. This also allows us to identify the color 
operator responsible for the breakdown of Reggeization at NLL 
for the imaginary part of the amplitude, and to determine the 
operator (\ref{NNLL}) that breaks Reggeization 
for the real part at~NNLL. 
We believe that our results, 
to be discussed in detail elsewhere, pave the way for further 
progress: corrections to the dipole formula may be further constrained; 
Reggeization of finite contributions to the amplitude may be studied from an infrared 
viewpoint; our results could be used to test the breakdown 
of Reggeization at NNLL and gauge its impact on phenomenology; finally, the infrared singularity structure of amplitudes may be used to study the high-energy limit beyond the realm of Reggeization, as well as other kinematic limits.

{\em Acknowledgments:\/}
We are grateful to G. Korchemsky for useful discussions. This work 
was supported by the Research Executive Agency (REA);  
by the LHCPhenoNet ITN, contract PITN-GA-2010-264564;
by MIUR (Italy), contract 2006020509$\_$004; by the STFC 
Fellowship ``Collider Physics at the LHC''; and by a SUPA 
Distinguished Visitor Fellowship (LM). CDW, CD, and VDD
thank the University of Edinburgh for hospitality.

\end{document}